\documentstyle[epsfig]{aipproc}

\begin{document}
\title{Nonthermal X-ray emission from young Supernova Remnants}

\author{Eric van der Swaluw Abraham Achterberg and Yves A. Gallant}
\address{Astronomical Institute, Utrecht University, The Netherlands}

\maketitle

\begin{abstract}
The cosmic-ray spectrum up to the knee ($E\sim 10^{15}$ eV) is attributed to 
acceleration processes taking place at the blastwaves which bound supernova 
remnants. 
Theoretical predictions give a similar estimate for the maximum energy which
can be reached at supernova remnant shocks by particle acceleration. Electrons 
with energies of the order 
$\sim 10^{15}$ eV should give a nonthermal X-ray component in young supernova
remnants. Recent observations of 
SN1006 and G347.3-0.5 confirm this prediction. 
We present a method which uses hydrodynamical simulations to describe the
evolution of a young remnant. These results are combined with an algorithm 
which
simultaneously calculates the associated particle acceleration. We use 
the test particle approximation, which means that the back-reaction on
the dynamics of the remnant by the energetic particles is neglected.
We present synchrotron maps in the X-ray domain, and 
present spectra of the energies of the electrons
in the supernova remnant. Some of our results can be compared directly with 
earlier semi-analytical work on this subject by Reynolds [1].
\end{abstract}
\section*{Introduction}
The number of supernova remnants (SNRs) with an observed nonthermal X-ray 
component is slowly increasing (Allen et al.[2]). This confirms the theory 
of diffusive shock acceleration
(DSA), which predicts the production of relativistic particles at SNR shocks 
in the required energy range. Furthermore it strengthens the evidence that
SNR shocks are indeed the sites where cosmic rays are accelerated up to the 
knee of the observed cosmic ray spectrum ($E\sim 10^{15}$ eV). 
It is expected that with the current observational facilities like Chandra and 
XMM, the details about SNRs with a nonthermal component and, hopefully, the 
number of SNRs identified as sources of nonthermal X-ray emission will increase.
This motivates the extension of the
current models of nonthermal X-ray emission of SNRs, in order to keep up  with
with the status of observational work on this subject. We present a model which
consists of a hydrodynamics code calculating the flow of an evolving SNR coupled
with  
an algorithm which simulates simultaneously the transport and acceleration of 
relativistic particles. 

\section*{Hydrodynamics}

The evolution of a single supernova remnant (SNR) can be divided in four 
different stages (Woltjer [3]): the free expansion stage, the Sedov-Taylor 
stage, the pressure-driven snowplow stage and the momentum-conserving stage. 
We will only focus on the first two stages. In later stages, both 
synchrotron losses and the efficiency of the acceleration process, prevent 
particles to be accelerated to energies where they can emit X-rays by the
synchrotron mechanism. In the free expansion
stage the material of the SNR is dominated by expanding ejecta from the 
progenitor star, bounded by a shock which is driven into the interstellar medium
(ISM). Due to deceleration 
by the ISM, a reverse shock accompanies the forward shock (McKee [4]). When the 
SNR has swept up a few times the ejected mass of the progenitor star, this 
reverse shock is driven back into the interior of the SNR. This marks the stage
where the
expansion of the SNR shock will make the transition to the Sedov-Taylor stage. 
We have performed 
hydrodynamical simulations for a young supernova remnant up to this transition.
All the hydrodynamical simulations were performed with the 
Versatile Advection Code \footnote{See http://www.phys.uu.nl/\~{}toth/} 
(VAC, T\'oth [5]). 
The calculations are 
performed
on a spherically symmetric, 1D, uniform grid. As an initial condition we deposit
thermal energy and mass in the first few grid cells. This leads to the formation
of both the reverse shock and the forward shock, discussed above. The grid
resolution is such that the forward and the reverse shock are both 
resolved, and give the right compression factor ($r=4$) for a 
non-relativistic strong shock.
We make the model of the SNR 2D axially symmetric by imposing
a uniform magnetic field in the ISM, aligned with the symmetry axis, at the
start of the simulation. By using the condition of a frozen-in 
magnetic field (ideal MHD) the magnetic field lines are dragged along with 
the fluid, determining the configuration of
the magnetic field at later times. 

\section*{Particle Acceleration}

In principle, the acceleration and propagation of particles in a flow can be 
investigated by solving a Fokker-Planck equation for the particle distribution
in phase space (e.g. Skilling [6]). Instead, we employ a method 
which uses It\^o stochastic differential equations (SDEs). The SDE method
simulates the random walk trajectory of a test particle in a given flow. By
considering many realizations of the SDE in the same flow one constructs the 
distribution of particles in phase space. This corresponds to the solution 
of the Fokker-Planck equation (Achterberg and Kr\"ulls [7]).

By performing hydrodynamical simulations of a young SNR, and by using the flow 
from these simulations as an input for the SDE method, we simultaneously 
describe {\it both} the particle acceleration and the hydrodynamical evolution.
Losses due to synchrotron radiation and inverse Compton radiation 
are easily included in the SDE method. In this way, we get the 
electron distribution in phase space for a young SNR which realistically 
describes: 
(1) acceleration at the forward shock, (2) adiabatic losses due 
to the expansion of the SNR, (3) synchrotron losses due to the presence of 
magnetic fields and (4) inverse Compton losses due to the interaction of the 
electrons with the photons of the cosmic microwave background.

\begin{figure} 
\centerline{\epsfig{file=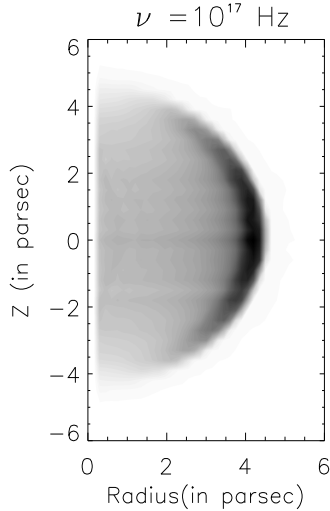,height=2.8in,width=3.8in}}
\vspace{10pt}
\caption{Synchrotron map for $\nu = 10^{17}$ Hz.}
\label{fig1}
\end{figure}

\section*{Results}
We present the results from a simulation of a SNR with a total 
mechanical energy of $E_0\simeq 0.93\times 10^{51}$ erg and an ejected mass of 
$M_{\rm ej} = 5.6 M_\odot$. This expands into an uniform medium with density 
$\rho_0 = 10^{-24}$ g/cm$^3$ and an axially symmetric magnetic field
with strength $B_0 = 10 \mu$G. 
We continuously inject particles at the forward shock of
the SNR, starting at an age of $t=100$ years up to the end of the simulation at
an age of $t=1000$ years. A
total of
$\sim 3.7\times 10^6$ test particles were injected during the simulation. The
injection was at a fixed momentum, and proportional with the area of the
remnant, with a particle weight which
takes into account the increase in the amount of material processed by the shock
as it expands. 
\begin{figure} 
\centerline{\epsfig{file=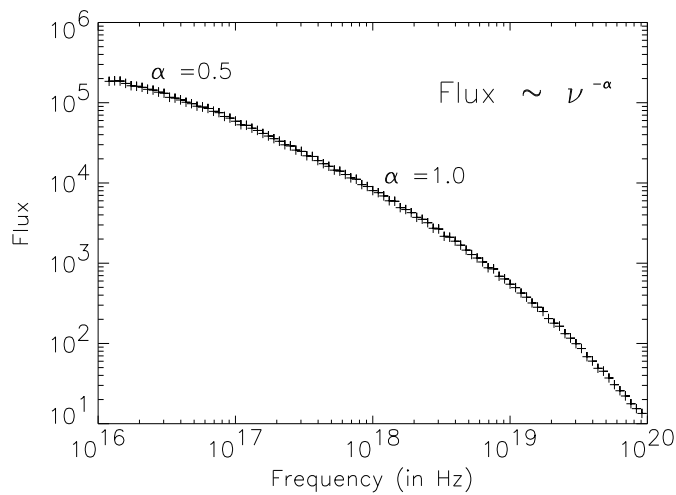,height=2.8in,width=3.8in}}
\vspace{10pt}
\caption{Spectrum of the total remnant.}
\label{fig1}
\end{figure}
At the end of the simulation we have the position and the momentum of each
particle in the simulation box. Because the magnetic field strength
throughout this box is also known, we can produce synchrotron maps at different frequencies. One
example is shown in figure 1, which is reminiscent of the synchrotron maps
presented by Reynolds [1]. Furthermore, figure 2 shows the spectrum of all the 
particles in the remnant. At low frequencies one can 
see the expected value of the spectral index for acceleration at 
a strong shock (compression ratio $r=4$), i.e. $S_{\nu}\propto \nu^{-0.5}$. 
In the roll-off part of the spectrum, where the synchrotron and inverse Compton 
losses start to 
compete with the energy gain due to the acceleration process, we get a 
power-law index of $S_{\nu} \propto \nu^{-1.0}$ at a frequency of $10^{18}$ Hz.
\section*{Conclusion}
We have presented results from a method using a combination of
hydrodynamical simulations and an algorithm simultaneously simulating
particle acceleration in a SNR. The results are comparable with earlier work 
on this subject by Reynolds[1]. Future work will consider the evolution of a
SNR in an ISM which is not uniform, like stellar wind cavities or molecular 
clouds.
The combination of the SDE method and hydrodynamics is a strong
tool to investigate the resulting morphology of the synchrotron emission.

\end{document}